\documentclass[english]{article}
\usepackage[auth-lg]{authblk}
\usepackage[T1]{fontenc}
\usepackage[utf8]{inputenc}
\usepackage{textcomp}
\usepackage{amsthm}
\usepackage{amsmath}
\usepackage{graphicx}
\usepackage{amssymb}
\usepackage{esint}

\makeatletter

\newcommand{\lyxmathsym}[1]{\ifmmode\begingroup\def\b@ld{bold}
  \text{\ifx\math@version\b@ld\bfseries\fi#1}\endgroup\else#1\fi}

\theoremstyle{plain}

\makeatother

\usepackage{babel}

\begin{document}

\title{Statistical mechanics formulation of radiobiology}

\author[1]{O. Sotolongo-Grau}
\author[1]{D. Rodriguez-Perez}
\author[2]{J. A. Santos-Miranda}
\author[1,3]{J. C. Antoranz}
\author[3]{Oscar Sotolongo-Costa}
\affil[1]{\small{Departamento de Física Matemática y de Fluidos, Universidad Nacional de Educación a Distancia (UNED), Spain}}
\affil[2]{\small{Servicio de Oncología Radioterápica, Hospital General Gregorio Marañón, Spain}}
\affil[3]{\small{Cátedra de Sistemas Complejos “Henri Poincaré”, Universidad de La Habana, Cuba}}
\date{~}

\maketitle

\begin{abstract}
The expression of survival factors for radiation damaged cells is
empirical and based on probabilistic assumptions. We obtain it either
from the maximum entropy principle for the classical Boltzmann-Gibbs
entropy and/or from the Tsallis entropy. Empiric models are found to be
particular cases of the obtained expression. The survival factor exhibits
a phase transition behaviour. This formulation supports different
tissues grouped as universality classes.
\end{abstract}
Radiobiology, Survival factor, Entropy, Statistical mechanics

\section{Introduction}

Empiric laws are at the grounds of science. However, when a scientific
discipline grows it is possible that the initially developed empiric
laws become unable to explain all new experimental data obtained.
Then engineering corrections and fitting coefficients emerge and usually
explain those unexplained behaviors. When the experimental
data grows enough even those corrections cannot explain the results;
then a new set of corrections is looked for, found and a proper explanation
for the new observed phenomena is provided. Eventually a large amount
of anomalous behaviours is observed and the empiric law reaches its
limit. It is impossible to go any further without getting deeper. 

At this point, a theoretical approach able to successfully explain
the data is needed. Even if the new theory brings too complex expressions
(like in the theory of relativity), the limits of every approximation
will be clear and it will be possible to reformulate the entire experimental
approach.

The tissue effect is an empirical concept widely used by oncologists
to find, given a radiation dosage, the survival factor of tissue or
tumor cells \cite{STEEL}. It is widely used in clinical radiotherapy
for both tumor and tissue cells and can be used to predict the outcome
of a treatment. It has been developed for years and it is capable,
as a concept, of gathering together several models of interaction
between cells and ionizing radiation. Even when using the {}``single
hit, single target'' , the {}``single hit, multi-target'', the
{}``two hit, single target'' or the linear-quadratic model \cite{Tubiana}
the powerful principle of grouping lethal events on a single adimensional
quantity remains. 

The general expression used in radiobiology for fraction of survival
cells is deduced assuming that lethal events per cell follow a Poisson
distribution \cite{Tubiana}. We think, besides, that the dependence
between survival factor and tissue effect could be found from deeper
laws based on first principles of nature.

The maximum entropy principle is completely universal, well established
and has an almost unlimited range of applications in physics, biology,
demography, economy etc. In its modern formulation it says, “Given
a model of probability distributions, choose the distribution with
highest entropy.” \cite{Harremoes01}. It means that one should look
for a distribution, consistent with the observed constraints, which
maximizes the entropy.

\section{Radiobiology and extensivity}

According to the radiobiology linear model for the cell survival factor,
the fraction of tissue killed by a radiation dose $D$ is\begin{equation}
F_{d}=1-exp[-\alpha D]\label{eq:fd}\end{equation}

Then $F_{s}=1-F_{d}$ is the cumulative probability of cell survival.
From this follows that\begin{equation}
p\left(D\right)dD=\alpha exp[-\alpha D]dD\label{eq:pD1}\end{equation}
is the killed cell probability density (per dose unit). The cumulative
probability fulfils the additive property:
\begin{equation}
F_{s}\left[D_{1}+D_{2}\right]=F_{s}\left[D_{1}\right]F_{s}\left[D_{2}\right]\label{eq:cumFD}\end{equation}
meaning that the effects of radiation are cumulative following an
additive model. The additive variable here is the tissue effect, $E=\alpha D=-\log\left[F_{s}\right]$
whose sum is

\begin{equation}
E\left[D_{1}+D_{2}\right]=E\left[D_{1}\right]+E\left[D_{2}\right]\label{eq:cumE}\end{equation}

However, the linear model is not accurate enough for a higher radiation
dosage and the empiric experience shows that the tissue effect must
be calculated as\begin{equation}
E=\alpha D+\beta D\lyxmathsym{\texttwosuperior}\label{eq:elq}\end{equation}
in what is called the LQ Model. The main problem involved here is
that the probability function is not extensive. That is, the survival
fraction for two doses is,

\begin{equation}
F_{S}[D_{1}+D_{2}]=\exp[-\alpha(D_{1}+D_{2})-\beta(D_{1}^{2}+D_{2}^{2}+2D_{1}D_{2})]\label{eq:F12}\end{equation}
and this implies,\begin{equation}
F_{S}[D_{1}+D_{2}]=F_{S}[D_{1}]\cdot F_{S}[D_{2}]\cdot\exp[2\beta D_{1}D_{2}]<F_{S}[D_{1}]\cdot F_{S}[D_{2}]\label{eq:F12des}\end{equation}
This means that the survival fraction is lower if the radiation is
applied in a single higher dose than if the same amount of radiation
is applied in two doses. Or, in other words, a continuous radiation
application kills more cells than a fractioned one. As $E=-\log F_{S}$
we can arrive to \begin{equation}
E[D_{1}+D_{2}]>E[D_{1}]+E[D_{2}]\label{eq:Edes}\end{equation}
meaning that $E$ is a nonadditive variable. As a result of the nonlinear
nature of $E$ in this case, the superposition principle is not fulfilled.

\section{The classical approach}

Prior to reproduce the Tsallis entropy discussion for the nonextensive
model, we will study the linear model extensive problem. This model
fulfils that if the amount of absorbed radiation is unlimited, no
single cell will survive,

\begin{equation}
F_{s}[D\rightarrow\infty]=0\label{eq:f-inf}\end{equation}

If the tissue effect is defined as proportional to the absorbed radiation
$E=\alpha_{0}D$ and $p(E)$ is the probability density of killing
a cell, then the fraction of killed cells for a tissue effect lower
than $E$ will be \begin{equation}
F_{d}(E)=\int_{0}^{E}p(x)dx\label{eq:fd_bg_def}\end{equation}
and the survival fraction for every tissue effect lower than $E$
will be\begin{equation}
F_{s}(E)=\int_{E}^{\infty}p(x)dx\label{eq:fs_bg_def}\end{equation}
Using $p(E)$ it is possible to write a Boltzmann entropy functional,
defined as

\begin{equation}
S=\int_{0}^{\infty}dE\, p(E)\log\frac{1}{p(E)}\label{eq:shannon}\end{equation}

Under the maximum entropy principle the expression of $p(E)$ can
be found if some assumptions are established. The completeness principle,\begin{equation}
\int_{0}^{\infty}p(E)dE=1\label{eq:normBG}\end{equation}
and the mean value existence,\begin{equation}
\int_{0}^{\infty}p(E)EdE=\left\langle E\right\rangle <\infty\label{eq:meanBG}\end{equation}
are demanded. 

Using the maximum entropy principle, the functional \begin{equation}
\int_{0}^{\infty}p(E)\log\frac{1}{p(E)}dE+a\int_{0}^{\infty}p(E)dE+b\int_{0}^{\infty}Ep(E)dE\label{eq:shannon_funct}\end{equation}
is built. Here $a$ and $b$ are the Lagrange multipliers and can
be found following the imposed restrictions and maximizing the functional.
It is straightforwardly obtained\begin{equation}
p(E)=\frac{1}{\left\langle E\right\rangle }e^{-\frac{E}{\left\langle E\right\rangle }}\label{eq:BGE}\end{equation}
the exponential distribution for the tissue effect.

The survival probability of a single cell will be\begin{equation}
F_{s}=e^{-\frac{E}{\left\langle E\right\rangle }}\label{eq:psBG}\end{equation}

We must note that \eqref{eq:psBG} is the experimentally proved and
the normally used expression for the survival factor as a function
of tissue effect and justified in the literature only through probabilistic
arguments \cite{Tubiana}. We can take $\alpha=\alpha_{0}/\left\langle E\right\rangle =1/\left\langle D\right\rangle $
and the expression \eqref{eq:psBG} gets written in the known standard
radiobiology form.

\section{The generalized approach}

Now that the extensive problem is solved, we will look for the non
extensive solution in a similar way. To apply this principle we demand
that there exists some amount of absorbed radiation $\Delta<\infty$,
or its equivalent tissue effect, $\Omega=\alpha_{0}\Delta$, after
which no cell survives, \begin{equation}
F_{s}[\Omega]=0\label{eq:cond1}\end{equation}
We will propose the use of Tsallis entropy \cite{Tsallis98}, \begin{equation}
S_{q}=\frac{1}{q-1}\left(1-\int_{0}^{\infty}p^{q}(E)dE\right)=\frac{1}{q-1}\left(1-\int_{0}^{\Omega}p^{q}(E)dE\right)\label{eq:TS}\end{equation}
as the generalized entropy. We impose the conditions \begin{equation}
\int_{0}^{\infty}p(E)dE=\int_{0}^{\Omega}p(E)dE=1\label{eq:normT}\end{equation}
and 

\begin{equation}
\int_{0}^{\infty}p^{q}(E)EdE=\int_{0}^{\Omega}p^{q}(E)EdE=\left\langle E\right\rangle _{q}<\infty\label{eq:qmean}\end{equation}
and apply the method of Lagrange multipliers. The functional\begin{equation}
\frac{1-\int_{0}^{\Omega}p^{q}(E)dE}{1-q}+a_{q}\int_{0}^{\Omega}p(E)dE+b_{q}\int_{0}^{\Omega}dE\, p^{q}(E)E\label{eq:TsFunct}\end{equation}
is maximized under those conditions and the values of $\Omega$, $a_{q}$
and $b_{q}$ are found,\begin{equation}
\Omega=\frac{2-q}{1-q}\left(\frac{\left\langle E\right\rangle _{q}}{2-q}\right)^{\frac{1}{2-q}}\label{eq:omega}\end{equation}
\begin{equation}
a_{q}=-\frac{q}{1-q}\left(\frac{\left\langle E\right\rangle _{q}}{2-q}\right)^{\frac{1-q}{2-q}}\label{eq:aq}\end{equation}
\begin{equation}
b_{q}=-\frac{1}{2-q}\left(\frac{\left\langle E\right\rangle _{q}}{2-q}\right)^{-\frac{1}{2-q}}\label{eq:bq}\end{equation}

From where the probability density function \begin{equation}
p(E)=\left(\frac{2-q}{\left\langle E\right\rangle _{q}}\right)^{\frac{1}{2-q}}\left(1-\frac{1-q}{2-q}\left(\frac{2-q}{\left\langle E\right\rangle _{q}}\right)^{\frac{1}{2-q}}E\right)^{\frac{1}{1-q}}\label{eq:lmT}\end{equation}
is obtained. Then, the survival factor is \begin{equation}
F_{s}(E)=\left(1-\frac{1-q}{2-q}\left(\frac{2-q}{\left\langle E\right\rangle _{q}}\right)^{\frac{1}{2-q}}E\right)^{\frac{2-q}{1-q}}\label{eq:psT}\end{equation}
or using expression \eqref{eq:omega},\begin{equation}
F_{s}(E)=\left(1-\frac{E}{\Omega}\right)^{\frac{2-q}{1-q}}\label{eq:psD}\end{equation}
for every $E<\Omega$. From \eqref{eq:omega} it is not hard to see
that when we deal with the extensive limit ($q\rightarrow1$) then
$\Omega\rightarrow\infty$ as $\left\langle E\right\rangle _{q}\rightarrow\left\langle E\right\rangle $. 

Defining $E=\alpha_{0}D$ as in the previous subsection, we eventually
get the expression for the survival factor of cells under radiation,\begin{equation}
F_{s}(D)=\begin{cases}
\left(1-\frac{D}{\Delta}\right)^{\frac{2-q}{1-q}} & \forall D<\Delta\\
0 & \forall D\geqslant\Delta\end{cases}\label{eq:finalF}\end{equation}

Notice that whereas \eqref{eq:psD} is expressed as function of the
non measurable magnitudes $\Omega$ and $E$, \eqref{eq:finalF} expresses
the survival factor in terms of the measurable quantities $D$ and
$\Delta$. 

The value of $\Delta$ characterizes a critical point for cell survival
probability and divides the phase plane in two sections with very
different properties. For $D<\Delta$ probabilities of survival and
death coexists but when $D$ becomes equal to $\Delta$, a phase transition
takes place and no cell survives. This behavior resembles phase transition
in ferromagnetics near the Curie point. 

Rescaling the radiation dose
as $D/\Delta$ would allow to study the reaction of tissue cells under
radiation in a more general way.

\section{The $q\rightarrow1$ limit}

\begin{figure}
\begin{centering}
\includegraphics[scale=0.8]{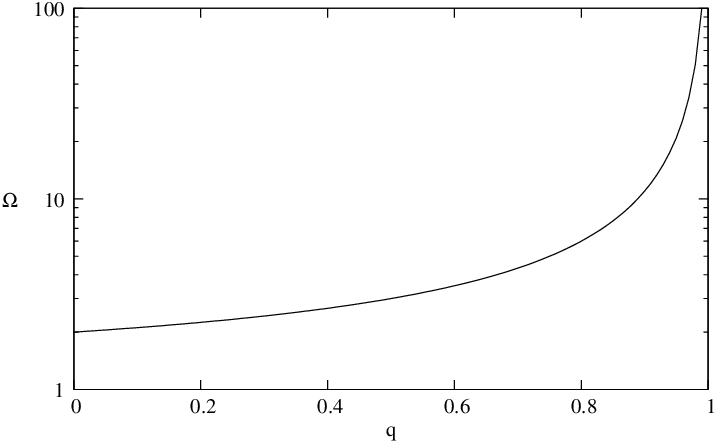}
\par\end{centering}

\caption{\label{fig:delta}Tissue effect limit ($\Omega$) as function of the
extensivity parameter ($q$) for $\left\langle E\right\rangle _{q}=1$. 
Even when in the plot $\left\langle E\right\rangle _{q}$
is assumed constant for every $q$, as the values of this parameter
remains enclosed, the general qualitative behaviour, near $q=1$,
must be close to the represented one. }

\end{figure}

The expression \eqref{eq:psT} must include the particular cases corresponding
to extensive systems. The cell survival probability limit ($\Omega$),
fits into this principle. Expression \eqref{eq:omega} gives the maximal
tissue effect as a function of the $q$-mean value of $E$. $\left\langle E\right\rangle _{q}$
remains bounded for any value of $q$. When $q$ tends to $1$, $\left\langle E\right\rangle _{q}$
tends to $\left\langle E\right\rangle $ and $\Omega$ diverges in
the extensive case. The obtained divergence is a trivial result that
raises from Boltzmann formulation. However for $q\neq1$ the divergence
disappears in agreement with the experience. As shown in figure \ref{fig:delta},
$\Omega$ is big only if the system is extensive enough. If $q$ is
far from the unity, every cell will be annihilated with a finite amount
of radiation.

If the system is almost extensive ($q\thickapprox1$), and the radiation
dosage is not too high, we can write the equation \eqref{eq:psT}
as the Taylor expression of the exponential function up to second
order,\begin{equation}
F_{s}(D)=\exp\left[-\alpha_{0}\left(\frac{2-q}{\left\langle E\right\rangle _{q}}\right)^{\frac{1}{2-q}}D-\frac{\alpha_{0}^{2}}{2}\frac{1-q}{2-q}\left(\frac{2-q}{\left\langle E\right\rangle _{q}}\right)^{\frac{2}{2-q}}D^{2}+O(D^{3})\right]\label{eq:finding_LQ}\end{equation}
and comparing with \eqref{eq:elq} we find\begin{equation}
\alpha=\alpha_{0}\left(\frac{2-q}{\left\langle E\right\rangle _{q}}\right)^{\frac{1}{2-q}}=\frac{1}{\Delta}\frac{2-q}{1-q}\label{eq:alpha_measure}\end{equation}
 \begin{equation}
\beta=\frac{\alpha_{0}^{2}}{2}\frac{1-q}{2-q}\left(\frac{2-q}{\left\langle E\right\rangle _{q}}\right)^{\frac{2}{2-q}}=\frac{1}{2\Delta^{2}}\frac{2-q}{1-q}\label{eq:beta_measure}\end{equation}

These expressions show that in the extensive case ($q=1$), then $\alpha=1/\left\langle D\right\rangle $
and $\beta=0$ recovering the linear model obtained from the Boltzmann
formulation. This also allows to establish a correspondence between
the known magnitudes $\alpha$ and $\beta$ from the widely used LQ
model and the newly defined parameters. Furthermore, the $\beta$
parameter of the LQ model can be interpreted as a second order approximation
of the non-extensive model. However, if $D$ is high enough or $\Delta$
becomes a small quantity, the LQ model becomes useless to describe
the survival factor.

\section{Comparison with experimental data}

To compare with experimental data we have selected some example plots
in \cite{STEEL} and \cite{HandbookRT}, where the survival factor
is plotted as a function of $D$ and separated curves are obtained
for different radiation conditions. We used \eqref{eq:finalF} to
fit. As it is a function of $D/\Delta$, all curves collapse to a
single one. This means that $\Delta$ is the natural unit of $D$.
Every plot of $F_{s}$ for a given tissue must converge to the same
curve if $D$ is expressed in the appropriate $\Delta$ units.

\begin{figure}[h]
\begin{centering}
\includegraphics[scale=0.6]{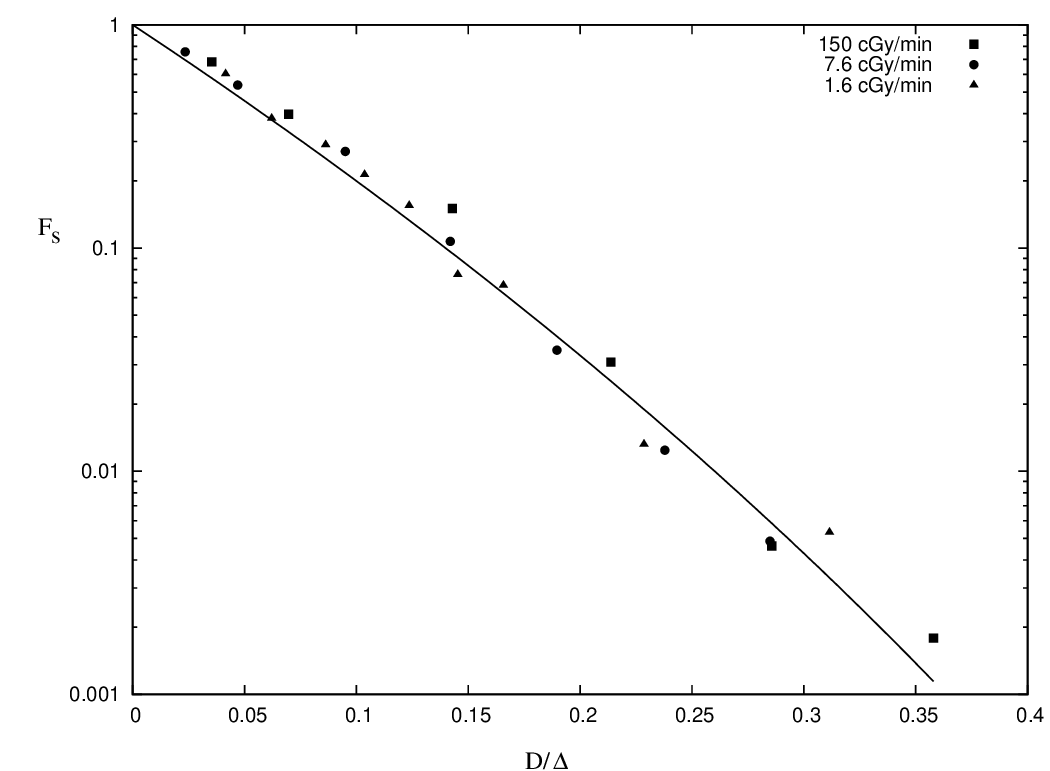}
\par\end{centering}

\caption{\label{fig:steel_dose_rate}Cell survival curves for a human melanoma
irradiated at dose rates of $150cGy/min$ ($\Delta=28Gy$), $7.6cGy/min$
($\Delta=42Gy$) and $1.6cGy/min$ ($\Delta=48Gy$). The solid line
is the survival factor as function of $D/\Delta$ for $q=0.93$. Data
was extracted from \cite{STEEL}. }

\end{figure}

Figure \ref{fig:steel_dose_rate} shows experimental data of a human
melanoma under radiation for $F_{s}$ as a function of $D/\Delta$
at different dose rates. 

We can see that, at least for the represented ones, even when the
dose rate affects the transition point $\Delta$, the exponent $\frac{2-q}{1-q}$
in \eqref{eq:finalF} remains constant. 

\begin{figure}[h]
\begin{centering}
\includegraphics[scale=0.6]{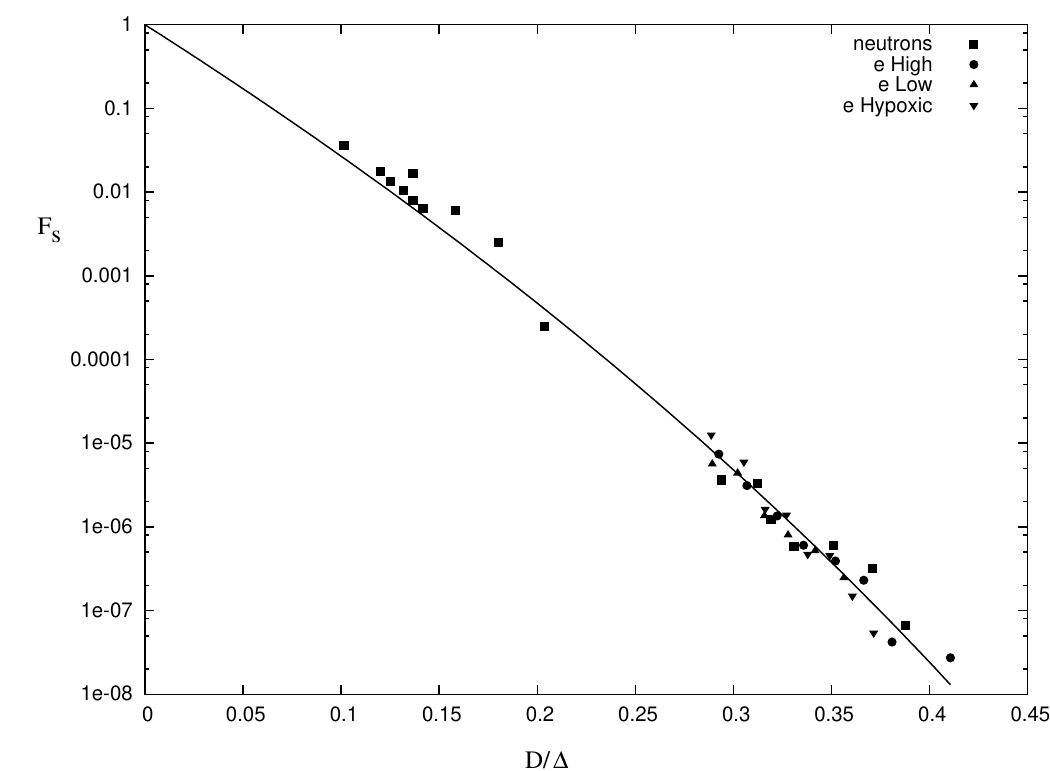}
\par\end{centering}

\caption{\label{fig:hb_haces}Survival curves for intestinal stem-cells. The
radiations were neutrons ($\Delta=39\mbox{ Gy}$), electrons {[}high
($\Delta=68\mbox{ Gy}$) and low dose rate ($\Delta=76\mbox{ Gy}$){]}
and electrons under hypoxic conditions ($\Delta=180\mbox{ Gy}$).
The solid line is the survival factor as a function of $D/\Delta$
for $q=0.97$. Data was extracted from \cite{HandbookRT}.}

\end{figure}

This is also true for the kind of radiation. Figure \ref{fig:hb_haces}
shows cell survival data, extracted from \cite{HandbookRT}, for
stem cells under a beam of neutrons and electrons in different conditions. 

Though the cell survival factor ($F_{s}$) as a function of $D$ depends
on multiple factors, like the dose rate or the kind of radiation,
the dependence of $F_{s}$ with $D/\Delta$ has a universal character.
In this case, as the variable $D$ is rescaled by $\Delta$, the curves
collapse and a universal behaviour emerges. The main factor then becomes
the exponent, dependent on $q$, which in fact divides the cell behaviour
under radiation in universality classes. The parallel with critical
phenomena becomes apparent.

\section{Conclusions}

We have found a theoretical approach that puts the well known LQ
model on physical grounds starting from first principles, rather than
on probabilistic assumptions attempting to express the survival factors
in terms of the tissue effect. Analyzing the used expressions we found
that the tissue effect is not additive if defined through the LQ
model. The non linearity of the tissue effect makes inapplicable the
superposition principle, i.e., the tissue effect due to a continuous
dose is higher than the corresponding to a fractioned one if the same
amount of total energy is applied. 

The survival factor expression as a function of the absorbed energy
per unit mass was derived from the maximum entropy principle following
the Boltzmann entropy. Since the linear model is obtained from the
Boltzmann entropy and it does not explain the known experimental data,
so, a more general approach is introduced. Using the Tsallis $q$-entropy
formulation a generalized expression for the survival factor is found.
The extensive limit shows that the obtained expression is consistent
with the known empiric laws. Besides, the empiric coefficients could
be expressed in the new terminology and a new interpretation of its
meaning is provided. 

The law obtained for the survival factor exhibits a phase transition
behavior similar to second order ferromagnetic phase transitions where
the imanated state corresponds to cell survival. A critical value
of the absorbed energy marks the frontier between the non survivor
cell region and the coexistence between survival and death probabilities.
The transition between both regions occurs with a non integer critical
exponent revealing a behavior similar to the ferromagnetic phase transition
at the Curie point. Rescaling every case with this coefficient allows
to find the common shape for different experimental data belonging
to the same tissue. When compared with available experimental data,
the phenomenon is shown as universal for a given tissue. Values of
$q$ would allow to group different tissues in universality classes.

\bibliographystyle{unsrt}

\pagebreak{}

\end{document}